\documentstyle[]{book}
\def\hang{\hangindent\parindent}
\def\textindent#1{\indent\llap{#1\enspace}\ignorespaces}
\def\litem{\par\hang\textindent}

\def\references{\leftskip15pt\parindent-15pt\footnotesize}

\def\edoc{\end{document}}

\begin{document}
\pagenumbering{arabic}
\addtocounter{page}{332}
\frenchspacing 
\parindent15pt 
\oddsidemargin 0mm
\evensidemargin 0mm

\renewcommand{\theequation}{\arabic{equation}}
\setcounter{equation}{0}
\pagestyle{plain}

\newcommand{\grs}{$\gamma$-rays }
\newcommand{\gr}{$\gamma$-ray }
\newcommand{\alg}{\alpha_{\gamma} }
\newcommand{\alp}{\alpha_{\rm p} }
\newcommand{\ale}{\alpha_{\rm e} }
\newcommand{\ep}{(e^{+},e^{-}) }
\newcommand{\alr}{\alpha_{\rm r} }
\newcommand{\alinj}{\alpha_{\rm inj} }
\newcommand{\alw}{\alpha_{\rm w} }

\leftline{}
\leftline{}
\leftline{}
\leftline{}

\Large
\centerline{\bf Workshop Summary}

\normalsize 

\leftline{}
\leftline{}

\large 
\centerline{Paolo Coppi}

\normalsize 

\leftline{}

\small 
\centerline{Yale University, Dept. of Astronomy, P.O. Box 208101, 
New Haven,}
\centerline{CT 06520-8101, USA}

\vspace{18mm}

\footnotesize 
\noindent 
{\bf Abstract.} I present a general overview of the results discussed
during the Cracow 1997 workshop on ``Relativistic Jets in
AGNs''. My emphasis will be on showing the significant progress
made in several areas over the last few years,
pointing out what I feel are some of the more important
issues still facing us today, 
and suggesting where progress is likely to
be made in the near future.

\vspace{15.5mm}

\normalsize 

\leftline{\large\bf 1. Introduction}
\vspace{5mm}

\noindent 
This workshop was unusual in that it brought together 
participants with a wide variety
of expertise, from observational radio astronomy to 
numerical modeling of acceleration processes in plasmas.
The rapid observational progress made in studying relativistic
jets  and the richness of the jet phenomenon
in general were evident in the impressive range 
of jet emission energies,
spanning over 15 decades
from Gigahertz radio frequencies to TeV gamma-ray energies,
and jet length scales, from astronomical units to megaparsecs,
discussed by the participants.

As a convenient reference point in time to highlight the progress
we have made, I~choose the end of the 
1980s when I was shipped off
by my graduate thesis adviser to take notes for him at a 
VLBI workshop in Socorro, New Mexico. With the exception of 
the enigmatic object SS 433 in our Galaxy, relativistic jets
seemed to me confined largely to galaxies belonging to the 3C 
radio catalog, in particular 3C 273, 3C 279, and 3C 345 which
were the subjects of numerous talks. At the time, VLBI ``experiments''
to detect fringes from sources were difficult undertakings
involving major international collaborations, hence the tendency
to look at the same sources. Plot after plot was shown
of highly distorted ``blobs'', many of which, despite
the distortions,  could be seen to move in roughly straight lines 
on the sky with apparent velocities exceeding the speed of light. 
Usually, but not always, the motion of the VLBI blobs was aligned
with the orientation of the jet on VLA scales. When it was not,
this was tentatively interpreted as an indication that the jet
was somehow being bent. The superluminal motion of the blobs, together 
with the lack of two-sided jets
on VLBI scales, and the fact that the fainter of the two VLA-scale radio jets
typically showed less polarization than the brighter one (presumably
because the faint one was oriented away from us and was
seen through more depolarizing matter) were taken as strong
arguments that we were seeing Doppler boosted emission from relativistically
moving fluid in the jets. Although the details were still being
debated at that meeting, the ``unified model'', 
where most differences
in radio morphology and jet power could be explained away as simply
a function of Doppler boosting and the observer's viewing angle,
seemed generally accepted.  The inner jets in powerful Fanaroff-Riley
II (FRII) sources seemed to have typical bulk Lorentz factors
$\sim 10,$ while those in  systematically less powerful Fanaroff-Riley I 
(FRI) sources had lower Lorentz factors $\sim 4.$ 
This conclusion,
however, was still based on superluminal motion studies of a rather small 
sample of objects.  This unification scheme, of course, applied only
to the ``radio loud'' quasars containing jets.  Most 
Active Galactic Nuceli (AGN), about 90\%,
were instead ``radio quiet'' down the milli-Jansky level
and did not show any jets.
\vspace{1pt}

\pagestyle{headings}
\markboth{Coppi}
{Workshop Summary}

The main radiation mechanism responsible for
the observed jet emission was thought to be synchrotron
self-Compton (SSC) emission from energetic electrons or 
electron-positron pairs. The Compton component of the emission
seemed to be more or less of a nuisance and upper limits on 
the Compton emission (at X-ray) energies were mainly useful
for putting lower limits on the bulk jet Lorentz factor (so
that the Compton catastrophe,  where all the SSC power ends up
in the Compton component, could be avoided). The emission
from jets was thus a phenomenon limited mainly to the radio and
optical, with a small component in X-rays. The COS-B 
satellite had detected
GeV gamma-ray emission from one jet source (3C273), but it was
weak and not thought to be associated with the jet. Balloon flights had detected strong MeV emission
in several other AGN that did not contain jets, e.g., 
NGC 4151. The GeV emission seen in 3C273 was probably just
the high-energy tail of a similar MeV emission component.
Indeed, at one point it was speculated that all AGN might have gamma-ray emission
at the level of the 3C273, and that this could explain the
gamma-ray background detected by Sas-II.) One exception 
to the view that jets should produce little high-energy
emission was the of paper Melia \& K\"onigl (1989) where
jets were presumed to start out with very high bulk Lorentz factors
($>10^3$) and radiatively decelerate to their observed terminal
Lorentz factors $\sim 10.$  At that workshop,
Arno Witzel also gave one of his first talks in which he swore 
he had observed the radio emission from one source vary 
$\sim$ 20\% on a few hour timescales. If intrinsic to the source,
this intraday variability (IDV) implied a very high brightness
temperature incompatible with the standard SSC models and
jet bulk Lorentz factors. People were interested, but no one
really knew what to make of this observation if it proved to
be correct. I also noticed one poster paper discussing a strange
new class of sources (``compact doubles'') where VLBI observations
of their centers showed two or three {\it stationary\/} emission
components which didn't fit into the standard relativistic
jet picture.  Overall, I came away from that meeting already thinking
that  relativistic jets were rather nifty, amazing objects. 
Little did I know what was to come, 
however. In the sections below, I will try to give an overview
of the exciting new results reported and reviewed during
the present workshop as well as of some of the still-outstanding problems
we need to address. I conclude by speculating on where some
of the advances in the next few years may come from. 
I apologize in advance for anyone's work I may have misrepresented
or left out; the errors and omissions\break are mine.

\vspace{6.5mm}
\leftline{\large\bf 2. Recent progress in understanding jets}
\vspace{4.5mm}

\leftline{\bf 2.1. Observations}
\vspace{3.5mm}

\leftline{\bf 2.1.1. VLBI (parsec-scale) radio jets in AGN}
\vspace{2.5mm}

\noindent 
Our VLBI observing capabilities have improved 
markedly over the last ten years. The maximum spatial resolution available
has increased due to our ability to observe at higher frequencies
and use longer baselines involving space satellites. The arrival
of the VLBA now lets us  make maps with
unprecedented dynamic range on a fairly routine basis, and
allows us to contemplate monitoring many more sources than 
was possible before.  One example of interest
is the observation of M87 by Junor \& Biretta 
(1995) which shows the jet in M87 exists and is apparently 
well-collimated down to distances $\sim$ 100 Schwarzchild
radii ($\sim 10^{16}$ cm) from the central black hole. Another
is the apparent detection of parsec scale {\it circular\/} polarization
at the 0.5\% level in 3C84 and 3C279 (Homan et al. 1997). If confirmed,
circular polarization will be an important diagnostic for the source
geometry and magnetic field structure and the energy
distribution of the emitting electrons (e.g., Bjornsson 1990).
One final one is the report by Gabuzda (this proceedings)
of the detection of rapid (intraday) variability in VLBI 
monitoring campaign of the BL Lac PKS 2155 where the polariztion
varied in one emission component but not another, which should
have implications for our understanding of the intraday variability
phenomenon. My impression, though, is that the benefits of these
new capabilities are just starting to be realized and more\break is to come.

Overall, the general picture emerging at that Socorro workshop seems 
to have withstood more detailed scrutiny. The inferred magnetic fields 
in the VLBI jets of powerful quasars still seem preferentially 
aligned parallel to the jet axis, while in BL Lacs, they are aligned
perpendicular to the jet axis. The current interpretation is that in
the BL Lac case we are seeing amplification of the transverse field 
by shock compression, and in the quasar case, we are seeing strong
shearing of the field. The important point is that this systematic
difference between quasars and BL Lacs seems
to be real (see the review of Gabuzda here). The phenomenon of 
misalignment between the small-scale VLBI and larger scale VLA
jets also seems relatively common and is interpreted as evidence
of jet bending, perhaps due to interaction with a surrounding
medium.
There is more evidence (e.g., Zensus et al. 1995) that
the jet on the VLBI scales can have a complicated structure, perhaps
due to bending or precession or simply due to a complicated 
shock structure in the jet (e.g., see contributions by Mart\'{\i} and
Gomez). Particularly in the object 3C345, 
the radio-emitting blobs appear to be shot out initially in different 
directions and then converge to move on the same trajectory in
the sky. Detailed studies of the blob motions in other objects
sometimes show deviations from 
straight-line motion, with blobs accelerating and decelerating.
The poster I had noticed at Socorro on ``compact doubles'' now seems to
have mushroomed into a full-blown field
of study of Gigahertz-Peaked Sources (GPS) and Compact Symmetric 
Objects (CSO) (see Bicknell here and Bicknell et al. 1997 
for an overview). These sources look like 
classical double radio-lobed sources except
that they are much smaller in scale (0.1 - 1kpc). 
The current thinking is that we are seeing the working surfaces 
of two-sided jets as they ram into a dense interstellar medium,
and that the observed radio spectrum at low frequencies is attenuated by 
free-free absorption and perhaps induced Compton scattering in the 
surrounding gas. Even on small scales, it thus appears that jet morphology 
{\it can\/} be significantly influenced by\break the environment.

\vspace{6mm}
\leftline{\bf 2.1.2. Interactions of jets with ambient matter on larger scales}
\vspace{4mm}

\noindent 
With HST we can now resolve optical features down to similar scales
as the VLA ($\sim$ kpc scales),  and we can begin to make detailed 
radio-optical comparisons. In his presentation, Falcke (this proceedings) 
showed HST pictures of an AGN Narrow Line Region (NLR) where much of the 
high excitation emission occurred on the edges of the radio
jet feature --- just what one might expect for a jet running into
matter, and perhaps shocking or entraining some of it.  In 3C 264
(Baum et al. 1997), HST sees an optical ring at a projected radius
of $300-400$ pc from the center of the source, which is likely
due to absorption by a dense circumnuclear gas disk seen 
nearly face-on. The corresponding Merlin (comparable to be VLA)
radio map shows an initially well-collimated jet that appears
to blow itself apart, losing its collimation and dimming considerably
just as it reaches the outer boundary of the HST ring ---  again what
one might expect if an initially relativistic jet ran into
and entrained dense gas. In a similar vein, 
Bicknell (this proceedings) argues that the low power jets seen
in some Seyferts are actually underluminous in radio and have much
higher kinetic power than one might at first suspect. The explanation
he proposes is that the jets are initially moderately relativistic
and then decelerate and stop radiating once they become mass-loaded
by entrainment. On even larger scales (100 kpc), we see evidence from 
a comparison of radio and ROSAT X-ray observations
that the jet NGC 1275 interacts with the surrounding intracluster
medium. Indeed, Bremer et al. (1997) argue that many of the properties
of powerful, high-redshift jet radio sources can be accounted for by
postulating the jets are embedded in strong cooling flows at the centers
of cluster. In sum, the evidence presented at this workshop and
elsewhere increasingly argues that jets do not exist
in isolation and that their morphology, radiative properties, and 
composition can be significantly altered by their interactions
with their local environments. To fully understand the jets we are 
seeing (e.g., to explain in part the distinction between
FRI and FRII sources), I would argue that we need to fully understand the 
jet-external medium interaction.
While this conclusion might not particularly suprise anyone,
I would also argue that only recently have we begun to seriously work on this
aspect of the jet problem (e.g., see Bicknell and Plewa in this 
proceedings).
\newpage

\vspace{6mm}
\leftline{\bf 2.1.3. Blurring the lines: radio jets in radio-quiet AGN}
\vspace{4.6mm}

\noindent 
For me, one of the more interesting results shown in this workshop
was the VLA detection by Falcke of a weak, but clearly jet-like feature in 
a ``radio-quiet'' Seyfert galaxy that according to our conventional
understanding should not show jets. This appears to be a rather
general result. When one looks hard enough, many  AGN show
evidence of jets or at least a flat spectrum radio core. 
A particularly striking example is the LINER (low luminosity
AGN) NGC 4258, that was shown by VLBI maser observations 
to contain a massive central object surrounded
by matter in a~beautifully Keplerian, cold disk. Using more VLBI
observations, Herrenstein et al. (1997) have shown the previously
known radio jet on parsec-kiloparsec scales in fact extends 
down to 0.015 parsec (2000 Schwarzchild radii) from the central
object.  It seems that the sharp distinction between radio-loud
and radio-quiet objects, at least as defined by the presence
of relativistic radio jets, is becoming increasingly fuzzy
(see Falcke, this proceedings). I discuss this more
in \S 3.6.

\vspace{6.5mm}
\leftline{\bf 2.1.4. High energy emission from relativistic jets}
\vspace{4.6mm}

\noindent 
Perhaps the most dramatic, recent occurence in our study of jets
is the discovery that relativistic jets are (very) strong high-energy
emitters. Not long after it was turned on, the EGRET gamma-ray detector
on GRO saw a huge flare from a position on the sky consistent with 
that of the blazar 3C 279. This turned out to be the tip of a large
iceberg. The detection was not  a fluke, and
to date, EGRET has detected almost 60 blazars extending to energies
$\sim 10$~GeV (See the contribution by von Montigny for an 
observational review of blazar emission).  
Ironically, the radio-quiet AGN like NGC 4151 that we expected
to have strong MeV gamma-ray emission turned out not to have any.
(The balloon detections seem to have been due to background 
subtraction problems.) The only previously known gamma-ray quasar 3C273
turned out not to be typical at all of AGN, and of the gamma-ray blazars
we know of today, 3C 273 turns out to be a rather feeble example.

The GeV flux levels observed from these blazars are quite impressive
--- too impressive in fact, corresponding to isotropic luminosities exceeding 
$10^{49}$ erg/sec in the brightest cases. If all quasars or even just radio
loud quasars (the parent population of blazars) emitted isotropically
in this way, then the sky would be glowing in GeV gamma-rays at levels
$\sim$ 10-100 above the observed ones. Also, if blazar
gamma-ray emission, like the rest of AGN emission, is ultimately tied to 
accretion  onto a black hole (presumably limited to a few times the Eddington
luminosity), we infer much higher central black hole masses for
blazars than we have come to expect from past AGN studies, e.g., of
their optical broad line luminosity. A simple way out of these problems
is to postulate that the emission we are seeing is strongly beamed.
This conclusion receives strong support from the detection of
strong flaring by GRO with doubling times less than a day, which
without applying beaming corrections, implies a rather small size
for the gamma-ray emission regions. Note that GeV gamma-rays
can photon-photon pair produce on X-rays, and 
that blazars are also strong, variable X-ray
emitters. If the observed X-rays and gamma-rays come from the same emission
region, then we should not be seeing the GeV gamma-rays we do (they
would all have pair produced) unless all the X-ray/gamma-ray 
emission we see is relativistically
beamed by Doppler factors at least $\sim 5-10$ and hence our naive
estimate of the emitting region's ``compactness'' (opacity to GeV
gamma-rays) is off by several orders of magnitude. Arguments
like these and the fact that strong GeV emission is only detected
from blazar AGN (radio loud quasars previously thought from 
radio-UV observations to have jets pointing in our direction)
clinched the association of the gamma-ray emission with 
the same relativistics jet inferred to exist from radio observations.

The surprises did not end with the launch of GRO, however. At about
the same time, the Whipple telescope (a ground-based, Cherenkov
gamma-ray detector) significantly improved its sensitivity and 
suddenly detected the weak, nearby BL Lac source Mkn 421 at 
TeV (!) energies. This source and at least one other, Mkn 501, have since
been confirmed by other detectors as TeV emitters. The detection
of TeV emission has cosmological implications since TeV photons
from high-redshift source will pair produce on the cosmic infrared
background radiation before reaching us.
(See the 
contribution of Rhode for an overview of this and of TeV observations
in general.) 
As dramatic as the GeV variability seen by EGRET is, the TeV
variability apparently can be even more dramatic. In 1996,
Whipple detected several huge flares from Mkn 421 with
doubling times $\sim 30$ minutes (Gaidos et al. 1996),
and during this workshop, Mkn 501 was in similar flaring 
state, having increased its emission by a factor 
over 50 compared to that of the previous year.
In retrospect, there probably should have been more definite
predictions of strong gamma-ray emission from blazars. Radio
to X-ray observations of jet synchrotron emission had already
shown jets to contain very energetic electrons, and we also
knew that many of the radio loud quasars showed strong 
optical emission originating from their accretion disks, i.e., they had
very intense, unbeamed, radiation fields in their centers. If some
of the energetic jet electrons resided near the central radiation field,
a huge ``Compton catastrophe'' would result, producing copious
GeV-TeV emission (Near the central black hole,
Compton cooling of energetic jet electrons by the
``external'' unbeamed radiation dominates strongly over both 
synchrotron cooling in the jet magnetic field and Compton cooling
on synchrotron photons produced in the jet). 

In terms of understanding jet physics, the detection of gamma-ray
emission is important as it lets us measure exactly how much
 radiative dissipation is occuring in the jet (Before
GRO, theorists could invoke arbitrary dissipation as long
it occurred at unobservable gamma-ray energies. With GeV measurements
of individual sources and of the diffuse gamma-ray background, that window of
escape is now closed. Unless most of the jet power is in a component
that does not couple electromagnetically like neutrinos, which
is unlikely, we know  the bolometric luminosities of jets
to within factors of a few. Note that one cannot hide power at
higher photon energies as it will cascade down to GeV energies
and overproduce the background there, e.g., Coppi \&~Aharonian
1997.). As in the case of pulsars, the radio emission we have so laboriously
studied in AGN jets turns out to be a minor, energetically irrelevant, 
component of the total jet emission. In pulsars, the observed gamma-ray 
luminosity can often be $\sim 30\%$ of the total spin-down luminosity available
and often dwarfs the radio luminosity by two orders of magnitude.
In blazars, the gamma-ray power typically dominates the radio-optical 
synchrotron power by factors $\sim$ 10 and also appears to represent
a~significant fraction of the total jet kinetic power. Interestingly
the gamma-ray power, while large, does {\it not\/} appear to significantly
exceed the jet kinetic power inferred from studying the radio
lobes at the ends of the jets --- an important constraint on radiative
deceleration models for jets.

To make progress in understanding where and how jets dissipate
radiatively, we will need simultaneous, multi-wavelength observations
of blazar variability from radio to TeV energies (Except in
radio where we can do VLBI, we have little hope of spatially resolving
blazar X-ray/gamma-ray emission in the near future.). The results of 
recent monitoring campaigns were discussed in several talks at
the workshop see
in particular those of Maraschi and Takahashi). The current results
are far from being conclusive, but a general pattern seems to be 
emerging. Whenever a blazar is found in a high gamma-ray state, the
emission from $\sim$ optical to X-ray is also in a high state,
i.e., the emission at high and low energies seems to be strongly
correlated. When a sharp gamma-ray flare occurs, optical, UV, and X-ray
flares are also seen. Particularly in the BL Lac objects Mkn 421 and 
Mkn 501, the X-ray and gamma-ray (TeV) emission seem very tightly
correlated, with little lag. Below X-ray energies, the amplitude
of the flaring seems generally to decrease, perhaps due to dilution from
a non-varying component, and my impression is that low-energy flaring tends
to lag the high-energy flaring (although I note that in one case,
an optical flare may have {\it preceded\/} a gamma-ray flare --- i.e., we
need many more observations). Determining which parts of
the spectrum lag or lead other parts of the spectrum is 
particularly crucial
to understanding the dynamics of the particles responsible for 
the emission and disentangling various emission components. 
Right now, it is not completely clear what we should expect, e.g., in
blazars where Compton upscattering of photons external
to the jet is important, a~flare can be caused by an increase in
either the number of target photons (optical leads gamma-rays) or 
the number of energetic particles (gamma-rays lead optical).
As an example of where we may already have made significant progress, 
Takahashi (this proceedings) presented a~set of 
ASCA observations of Mkn 421 that show the X-ray emission at
$\sim$  1 keV lagged that at $\sim 10$ keV during during a~strong
TeV flare. Also, when the X-ray spectral index was plotted versus
intensity at a~fixed observing energy, a~characteristic ``hysteresis'' curve 
appeared (see Takahashi). Both the lag and
hysteresis curve are exactly what one expects if the X-ray
emission is synchrotron and one is seeing the 
spectral response to a rapid injection of very energetic electrons
which then cool slowly to lower energies (e.g., 
see Kirk, this proceedings). This, together 
with the tight X-ray/TeV correlation, lends strong support to the 
the picture that, in BL Lacs at least, we are seeing SSC emission
(with the synchrotron component dominating at X-ray energies and
below, and the Compton component dominating at gamma-ray energies).
Although data like this is highly suggestive, I would like to stress
that it is still relatively sparse and the conclusions far from
certain.

The rapidly variable emission I have been discussing probably
comes from the inner, sub-parsec regions of the jet. However,
the outer parts of the jet almost certainly also emit at high energies, at 
least in X-rays. On the parsec scale, Unwin et al. (1997) have
shown that in 3C 345, the overall level of X-ray emission correlates
with the emergence of radio VLBI features. Worral (this proceedings)
showed evidence from ROSAT observations of centrally obscured
jets that there is extended jet emission, perhaps out to kiloparsec
scales, and Tashiro (this proceedings) presented positive ASCA detections
of X-ray emission from jet radio lobes, probably from
Compton upscattering of cosmic microwave background photons by the
energetic electrons in the lobes. If one is making blazar observations
without good spatial resolution and the source is not clearly
in a~very strong flare state, any interpretation should take
into account the possible contributions from the larger scale jet.

\vspace{6mm}
\leftline{\bf 2.1.5. The discovery of microquasars}
\vspace{4mm}

\noindent 
As a testament to the ubiquity of the jet phenomenon, 
the last few years have also seen the discovery
of jets from black holes right in our own
Galaxy (see the overview by Ziolkowski, this proceedings).
Although the masses of these black holes are eight
orders of magnitude smaller than those of powerful AGN,
the Galactic jets appear rather similar to the AGN ones, and in particular,
show superluminal motion.

\vspace{5mm}
\leftline{\bf 2.2. Phenomenology}
\vspace{3mm}

\leftline{\bf 2.2.1. General radio-loud quasar unification schemes}
\vspace{3mm}

\noindent 
The last ten years have seen significant progress in 
firming up the links between the various members of the zoo
of radio-loud AGN. The apparently diverse members of this zoo go by such 
names 
as X-ray Selected BL Lacs, Radio Selected BL Lacs, FR-I/II
radio galaxies, High Polarization Quasars, Optically Violent Variables,
Flat Spectrum Radio Quasars, Steep Spectrum Radio Quasars,
Gigahertz-peaked Sources, Compact Steep Spectrum sources --- to name
a few. The first crucial link between all these objects is that they
all appear to contain jets. The second is that all their jets
probably start out with moderately 
to very relativistic bulk velocities, i.e., the emission
from the inner jet will be Doppler boosted and anisotropic as a result. 
Depending
on the orientation of the jet relative to the observer, one will see
very different spectra.
Shastri (this proceedings) showed an example of this, 
where the ROSAT spectral index of radio quasars/galaxies varies
systematically with the core dominance parameter $R$ (an orientation
measure if jet emission is beamed). If the jet points towards
us, the boosted jet emission dominates and one sees an often
featureless, rapidly variable continuum. If the jet points away
from us, one sees only the more or less isotropic 
emission from the underlying black hole accretion disk (e.g, the
broad emission lines seen in radio-quiet AGN). 
As summarized by Urry in her talk (see the
review of Urry \&~Padovani 1995 for details), we have now have 
accumulated a fair body of evidence that a~gross unification
scheme, based on viewing angle and characteristic jet bulk
Lorentz factors as the main parameters, actually
works.  The remaining major differences, e.g., between the morphologies
of FR-I and FR-II galaxies, right now seem probably due to differences
in the initial jet power and the interaction of the jet with ambient
matter (FR-I jets are weak compared to FR-II jets and may entrain
considerable matter; Gigahertz-peaked sources generically show
high rotation measures  and depolarization, i.e., signatures
consistent with their being surrounded by dense~gas.).

\vspace{6mm}
\leftline{\bf 2.2.2. The ``strawman'' (SS+E)C model 
and blazar unification schemes}
\vspace{3mm}

\noindent 
At this workshop, a clear ``strawman'' model for explaining the
variety of blazar spectra we see (see the contributions of Ghisellini, 
Fossati, Kubo, Takahashi, and Takahara). Namely, that all blazar spectra
can be explained by a one-zone, homogeneous ``(SS+E)C'' model. In this
model, energetic electrons located somewhere near the central black hole
emit synchrotron photons, and these jet photons together with some
number of external photons are Compton upscattered by the electrons
to high energies. The only parameters for this model are the source 
region radius,
the bulk jet Lorentz factor, the energy of the external
radiation field in the source frame ($U_{\rm rad}$), the energy density of the 
magnetic field in the source frame ($U_B$), the minimum and 
maximum ``injection'' energies of 
accelerated electrons/pairs, the power law index
of the electron/pair energy injection function, and the total
power supplied to the injected pairs. 
I do not think this model can be right in detail,
e.g., cascading may be important in some sources and 
the emission we see is very likely the superposition of 
several components and/or electron acceleration events. However,
as a first order phenomenological model,
it seems to work remarkably well, 
especially for spectra during strong flares where one localized
emission component may well dominate. 
In a burst of enthusiasm,
Ghisellini (this proceedings) has fit all the broad-band blazar spectra
he could find and has come up with a very interesting correlation
hinted at by other participants:  the energy of the
electrons responsible for the emission at the
peaks of the synchrotron and Compton components seems to systematically
decrease as the sum of the magnetic and external radiation energy
density ($U_B+U_{\rm rad}$) in the jet frame increases.  For some reason,
perhaps a balance between acceleration and radiative cooling timescales,
powerful sources (which presumably have stronger
external radiation fields and jet magnetic field) do not accelerate
electrons to as high an energy as weak sources. This scenario
is appealing as it explains the puzzling
differences between X-Ray Selected BL Lacs and Radio Selected BL Lacs.
The Radio Selected BL Lacs are known to be more luminous on average,
hence their synchrotron emission should peak at lower energies
(in the optical-UV), and they would not be picked up as often 
in X-ray surveys. Also the scenario is consistent with the recent finding of Padovani et al. 
   (1997)  that the distribution of X-ray spectral indices for Flat Spectrum 
   Radio Quasars (powerful objects, several of which have been detected
   by EGRET at GeV energies) is consistent with that of LBLs (``Low-energy
   cut-off BL Lacs'', in their terminology) and with the Compton 
    component dominating the X-ray emission seen 
   in both these classes of objects. In this unified picture, then,
   the spectra of all blazars are essentially the same except that
   they are shifted up and down in frequency depending, roughly,
   on their luminosities (and probably details of exactly where the particle
   acceleration occurs). The complete picture may be 
   be more complicated, however, as the ratio of the luminosities
   in the Compton and synchrotron emission components 
   seems also to depend on the total source luminosity: in BL Lac objects, 
   the ratio never reaches values as high as those seen in quasars.
   This may reflect the fact that accretion disks in BL Lacs are 
   systematically underluminous compared to quasar ones, i.e., there
   are relatively fewer external photons in BL Lacs.

\vspace{5mm}
\leftline{\bf 2.3. Theory}
\vspace{2.5mm}

\noindent 
The preceding scenarios are very elegant, but they provide 
no answers as to why, for example, the ``(SS+E)C'' model should
have the physical parameters it does. For example, what 
determines the energy distribution of the relativistic 
electrons in jets?  Unfortunately, 
as noted by Li et al. (this proceedings),
``The need for understanding particle accleration is stressed by
every high energy photon we observe.'' 
As a theorist, I am slightly ashamed to
say we have not kept up with observations on questions like 
this. The theoretical questions posed ten years ago and the 
answers tentatively proposed for them are still largely the
same: What creates and collimates the jets in the first place? Does the black
hole play a role (e.g., via the Blandford-Znajek effect or 
the coupling of the black hole spin to the accretion disk)
other than providing a~strong gravitational potential? 
How are particles accelerated? What roles if any do shocks, MHD wave
turbulence, and large-scale ``parallel'' electric fields play?
(For discussions of these last points, see, respectively, the 
contributions by Kirk, Ostrowski, Li, and Colgate.) If 
the gamma-ray emission region, the place where the bulk of
the internal jet dissipation seems to occur, always lies 
in some preferred region of the jet, why? Is it the region
where a~Poynting flux-dominated jet is transformed into
a~particle-dominated one? (See the discussion by Levinson for
more.) Because the theoretical issues involved are rather technical, I 
will not discuss them further. Rather, I will try
to conclude on a more positive tone by noting that there
was evidence for significant theoretical progress at this workshop.
Relative to ten years ago, the theory of shock acceleration is in much
more robust shape. Also, we have developed much more powerful radiative
transfer codes and models (e.g., inhomogeneous, multi-zone
ones) that can be brought to bear once the 
observations tell us what direction to move in. Finally,
of particular relevance to 
radio observations, were the 
impressive jet simulations presented
by Mart\'{\i} and Plewa. The three-dimensional, high resolution 
   simulations of Mart\'{\i} were essentially
unthinkable  even five years ago. Today, instead,
we are beginning to carry out fully relativistic, 
three-dimensional, {\it MHD\/} simulations. Although the connection
between the jet fluid properties and particle acceleration (and radiation) 
is still lacking in these simulations, 
one can now begin to make more educated guesses
as to what a real jet should look like. The simulated VLBI observations
shown by Gomez were enlightening as they graphically illustrated the 
complicated emission patterns (e.g., see Lind \& Blandford 1985) that
can arise in a realistic fluid flow, where shocks and fluid elements 
temporarily move in different directions with different velocities.

\vspace{8.5mm}
\leftline{\large\bf 3. Some new and old unresolved problems}
\vspace{4.5mm}

\noindent 
As Meg Urry noted in her talk, ``The greater our sphere of knowledge,
the larger its surface of contact with our ignorance.'' Below, I 
summarize some of the unresolved issues brought up 
during the discussion section of the summary talk, plus a few 
others I feel are important.

\vspace{5mm}
\leftline{\bf 3.1. How variable is blazar emission?}
\vspace{3mm}

\noindent 
As discussed by Stefan Wagner, 
if one thing is clear about blazars, it is that their emission
is extremely variable at practically all wavelengths. 
At radio frequencies, we see
intraday variability in intensity and polarization
on the order of $\sim 20\%.$  
At GeV energies, EGRET has also
seen intraday variability, with the source PKS 1622$-$297 showing
an increase of a factor $\sim 4$ in less than 7 hours
(Mattox et al. 1997). 
In general, the longer EGRET has observed, the more extreme
the examples of GeV variability it has found and a 
structure function analysis indicates that the shortest
variability timescales have yet to be resolved (
e.g., Wagner, this proceedings).
If this were not enough, the new ground-based Cherenkov 
telescopes (Whipple \&~HEGRA)
have detected variability at $\sim 1$ TeV energies 
on $\sim$~half-hour 
timescales from Mkn 421 and Mkn 501 (Gaidos et al. 1996,
Bradbury et al. 1997, Catanese et al. 1997). 
Variability in the X-rays on comparable 
($\sim$ hour) timescales has also\break been seen.  

In light of this data, an obvious 
question that needs to be answered is what exactly is the shortest
variability timescale as a function of energy? The answer could
have a major impact on models. Already the 
observed variability timescales are embarassingly short. In the case of the radio intraday variability, implied
   source brightness temperatures as high as $10^{16-18}$ were
   were recorded in several cases, with the current record being
$10^{21}$ K (Kedziora-Chudczer et al. 1996) for  
$\sim$ 1 hour variations seen in PKS 0405$-$385. These values are
orders of magnitude higher than the standard inverse Compton brightness
temperature limit of $10^{12}$ K.  If this variability is intrinsic to the
source, i.e., it is not due to propagation effects such as 
microlensing or interstellar scintillation, then the bulk jet Lorentz
factor required to explain away the discrepancy in PKS 0405$-$385
is $\sim 60-1000$ depending on the exact emission geometry 
(e.g., see Begelman et al. 1994 who would argue for the value
of 1000). Such values are significantly higher than the typical Lorentz
factors derived, say,  from superluminal motion considerations and strong
beaming of this type is not compatible with the population statistics in 
unification schemes. In the case of PKS 0405$-$385, 
even if the variability is not completely intrinsic and is due
mainly to interstellar scintillation, 
the apparent brightness temperatures must still exceed $10^{16}$ K
(Blandford, private comm.), 
i.e. one still has a~problem. If such apparent high brightness temperatures
persist, we may be forced to consider alternate, coherent emission 
scenarios such as perhaps proposed in  Benford (1992). Coherent emission
is not without precedent in Nature, but the brightness temperatures
here are so high that is not clear how the radiation can escape
from the gas-filled nuclear region without significant attenuation
due to stimulated processes (Coppi et al. 1993, Blandford \& Levinson 1994).
That being said, the observations remain.
Another area where rapid fluctuations may lead to problems is 
the rapid ($\sim$~half-hour) X-ray and TeV flaring seen in Mkn 421/501. 
To explain such
rapid variability and at the same time avoid catastrophic 
pair production of the observed TeV gamma-rays on X-rays
from the jet, the flaring region must be smaller (and closer
to the central source?) than than previously 
thought, $\sim 10^{14}-10^{15}$ cm, and the emission must be beamed
by Lorentz factors $\sim 10-15$ (e.g., see Gaidos et al. 1996),
again higher than the typical Lorentz
factors $\sim 4$ expected from the both the superluminal motion observations
and beaming statistics of low-power BL Lacs like~these. 

On a more practical note, the details of this variability must
be observationally 
understood and theoretically accounted for if one hopes to ever
make realistic emission models for blazar jets. 
``Quasi-simultaneous'' X-ray and TeV
observations separated by order an hour are not really simultaneous
if sources like Mkn 501 vary by factors of a few on half-hour
timescales. A particularly critical quantity in models (see the 
next section) is the 
X-ray to gamma-ray spectral index, i.e., the amount of X-rays
produced for a given level of gamma-ray emission. The fact
that relatively few X-rays appear to be seen, for example, rules
out models with lots of cascading, where gamma-rays from the
jet pair produce before escaping the central region of the AGN.
In several cases, though, my impression is that the
conclusion that X-rays are ``few'' are based on comparisons
of an X-ray flux obtained with an integration of a few hours
versus a gamma-ray flux obtained with a typical integration
time of two weeks. Fits to such apparently, but not really,
simultaneous data can be potentially misleading. Also, in determining
the ``characteristic'' gamma-ray emission level and energetics
of blazars, we must remember that many of the  EGRET 
gamma-ray detections probably represent extreme, perhaps
atypical, flares in these sources. I would argue that we currently
do not have a good handle on what the quiescent or time-averaged gamma-ray
emission from blazars is (Essentially every blazar detected has
had its flux drop below the EGRET detection threshold.). 
A potentially sobering example of this comes from the 
recent results of Pohl et al. (1997). In order to ascertain what
contributions blazars make to the diffuse gamma-ray background
at GeV energies (note that the observed background represents an
integration over several years), the authors carefully coadd all the
photons detected by EGRET that are consistent with having come from
the direction of a strongly detected blazar. The resulting
time-averaged, composite spectrum is rather soft (photon number index
$>$ 2) and does not look much like either the gamma-ray background
spectrum nor the hard flare spectra of blazars. See the 
contribution of Magdziarz, Moderski, \& Madejski for more
discussion of some of the issues connected with 
gamma-ray variability.

\vspace{6mm}
\leftline{\bf 3.2. What are jets made of and the location}
\leftline{\bf\phantom{3.2.} of the gamma-ray emission region}
\vspace{3mm}

\noindent 
This general topic is one that has plagued theorists
for years and received considerable discussion at
this workshop. The new observational data we have available
give us some important constraints, but the issue is 
far from resolved. As the relevant arguments
are summarized well in the contributions from Celotti and
Levinson, I  will only repeat the 
highlights here. A popular explanation for observations
like that of low Faraday polarization in AGN jets has been that 
jets are made of electron-positron pairs.  However, the detection  
of strong gamma-ray emission from blazar jets appears to rule
out scenarios where the bulk of the jet energy near the 
black hole is carried by pairs. The central radiation field
in an AGN is typically very intense and the radiation
drag on pairs is correspondingly large. At distances less than 
$\sim 10^{16}$ cm, pairs (and the jet itself if it is pair-dominated)
will be decelerated to Lorentz factors of a few. Comparing the
total radiative output of jets (typically dominated by 
their gamma-ray emission)
with the kinetic jet power inferred from the radio lobes at the 
ends of the jets, we are finding that the radiative power of
the jets can be comparable to, but not significantly greater than
the kinetic power. Thus, ``bulk'' jet deceleration
scenarios like that of Melia \& K\"onigl (1989) (where
an initially very fast jet is decelerated to a~terminal
Lorentz factor $\sim 10$) appeared to be ruled out. The inner
jet cannot suffer catastrophic radiative losses, and thus
if it is dominated by pairs, it must be cold
and slow --- often slower than the Lorentz factors $\sim 10$
inferred for the parsec scale jets.  The jet
must be slowly accelerated to its terminal Lorentz factor,
in which case the jet power is initially in some other 
form, e.g., Poynting flux or protons which do not radiate efficiently.
\vspace{1pt}

Leaving aside temporarily the issue of the  form in which the
bulk of the jet energy resides, 
another issue is where and how the pairs in the jet would actually
be produced.  In strong sources like 3C279, if the pairs are produced 
near the center where the jet particle density is presumably
high, they will annihilate away before propagating to the parsec scale
where we can use VLBI observations to constrain the actual 
density of pairs. Producing pairs too far away from the sources
is problematic, however, because the most efficient way to 
produce pairs is by photon-photon pair production, which requires
an intense field of target photons.   Pair production off 
 internal jet photons
is in principle possible, but the result is a compact ``fireball''
of the type discussed here by Thompson which could explain emission from
blazars which show a strong spectral cutoff above an 
MeV, but not from the bulk of blazars which appear to have emission
extending to GeV energies and beyond.
Hence, the best site for producing pairs is near the center
of the AGN, where the external radiation field is 
the strongest. The target photons of interest are probably
UV/X-ray photons. From kinematics  considerations, this means 
the pairs they produce must have energies above at 
least $\sim 10$ MeV. Since the pair production and 
Compton scattering cross-sections are of comparable magnitude,
if pair production is efficient, then Compton cooling is efficient
and the pairs that are produced will  Compton upscatter
ambient photons to X-ray energies and cool (barring some unforseen 
heating mechanism for the pairs). Because of the observationanal
constraints on the X-ray flux of blazars, the total number of pairs that can
be produced is thus significantly constrained. 
In addition, since the pairs cool, they carry away very
litle of their initial energy. Moreover, if there are too many
of these cooled pairs, Comptonization
by the pairs (moving with the bulk Lorentz factor of the
jet) should produce an observable excess of emission at soft X-ray energies,
the so-called Sikora ``bump'' which has not been observed yet.
While jets at parsec scales and beyond may contain significant numbers of 
pairs, because of arguments like these, it seems to me unlikely that
in {\it strong\/} sources the bulk of the jet energy is carried
by pairs at large distances (In weak FR-I/BL Lac
sources, many of the preceding arguments break down and
the jets could well be dominated by pairs.). However, I~note that
this conclusion is not the conventional one, and it faces a 
possibly significant
problem of energetics. Takahara (this proceedings) used his emission
model to estimate the jet kinetic power in 3C279. If the
observed SSC radiation comes from electrons neutralized by
ambient protons instead of from pairs, the inferred  jet
power increases from $\sim 10^{46}$ erg/sec to $10^{48}$ 
erg/sec, which starts to be uncomfortably~large. 

The last estimate depends critically on the lowest energies
of the pairs in the SSC emission region. The higher the minimum
electron Lorentz factor, $\gamma_{\rm min},$ the lower the number
of electrons in the source, the lower the number of required
protons and thus the lower the jet kinetic power. If electrons
could somehow be maintained at high Lorentz factors
($\gamma_{\rm min}
\sim 10$), this would solve not
 only the possible energetics problem
but also explain the low Faraday depolarization in the
absence of pairs since relativistic electrons effectively behave as 
heavier particles and induce less Faraday rotation. Although not
always noted, a minimum Lorentz
factor also enters crucially into the homogeneous SSC/external Compton model
for the observed spectrum. In order to match the observed spectral break at
$\sim$ MeV energies, Ghisellini (this proceedings) for example assumes 
that energetic electrons are injected into the radiation zone with a power
law energy distribution of cut off at a minimum  Lorentz factor
$\gamma_b \sim 100$. 
The steady-state Compton upscattered spectrum given such an 
electron injection function is a broken power law with energy spectral
index $\sim 0.5$ below a few MeV and a steeper spectral index above
(determined by the index of the electron injection function). The spectrum
above 1 Mev can be made arbitrarily steep, giving a change $\Delta 
\alpha_{X-\gamma} > 0.5,$ as observed in some cases (Without a minimum
energy injection energy and contrived cooling rates, inefficient cooling
of low energy pairs, the other mechanism proposed for MeV break,
can only produce $\Delta \alpha_{X-\gamma}=0.5.$). Why such a minimum injection
Lorentz factor $\sim 100$ should exist with this value is
an open question. 
Perhaps an important clue comes from the anti-correlation shown
here by Ghisellini between $\gamma_b$ and the total magnetic 
plus radiation energy density in the source region.
I note that cascade models, e.g., those discussed by Levinson,
naturally predict such a $\gamma_b$ (it is the minimum energy of the
produced pairs), but I have found such models generically
have problems explaining spectra like that of 3C273 where
$\Delta\alpha_{X-\gamma} > 0.5$ and the X-ray spectrum is hard
($\alpha_X=0.5$), i.e., it is not clear they apply in most objects.
\vspace{1pt}

Another still-open question, on which there was 
surprisingly little debate during this workshop (compared to 
others), is 
the exact location in the jet where energetic electrons/pairs are
accelerated and the observed gamma-rays are emitted. The consensus
(for strong sources)
seemed to be for distances $\sim 10^{16}-10^{17}$ cm from the 
central black hole. Cascade models typically can only work in
this range of radii because if most of the jet dissipation (electron
acceleration) occurs too close
to the black hole, too much X-ray flux is produced, and if the 
dissipation occurs too far, there are not enough target photons
to pair produce on and cascading is irrelevant. The external Comptonization models cannot work too close to the
    black hole because photon-photon pair production on disk photons
    will truncate the observed spectrum below a GeV, and they cannot
    work too far away (outside of the Broad Line Region) because the
    electron cooling times become too long and the characteristic source 
    sizes become too large to explain the rapid flaring seen by
EGRET. Purely SSC models 
are much less constrained since they provide their own seed photons,
but the typical source region sizes (variability timescales) and
magnetic fields used in models are characteristic of the subparsec
jet. Just because all these models agree roughly (to within
an order of magnitude) on where the gamma-rays come from is
still not definitive proof. In at least one object (Unwin et al. 1997),
some of the observed X-ray flux is clearly correlated with the
presence of particular VLBI ``blobs'' and is produced on 
parsec scales. The scenario of Mannheim (1993) where electron
acceleration and gamma-ray emission from shocks occur on the 
parsec scale should not be automatically dismissed, although I
feel it is more unlikely now  given the very intense, very 
rapid gamma-ray variability that has been seen. Also in two
cases (Wehrle et al. 1993, Pohl et al. 1995), strong gamma-ray
flares seemed to be roughly coincident with the time a VLBI
blob is extrapolated from its motion to have been emitted
from the origin of the jet, i.e., the gamma-ray emission
is associated with the formation of a blob, but it is over
before the blob is clearly distinguishable on the VLBI scale.
More attempts to correlate gamma-ray emission with the emergence
of blobs observed with high-resolution VLBI
would clearly be interesting. Even if we have correctly
guessed the location of gamma-ray emission region, another
question remains: {\it why\/} does so much dissipation of the
jet energy occur on these size scales? I don't currently
know of a very good answer (but see the contribution\break 
of Levinson).

\vspace{6.5mm}
\leftline{\bf 3.3. How many emission components do we need:}
\leftline{\bf\phantom{3.3.} is the one-zone emission model right?}
\vspace{4mm}

\noindent 
Occam's Razor suggests that we stick with the simple straw-man
model for blazar emission until the data requires otherwise.
With the current quality of data, the minute one allows for 
different source regions with different parameters, one 
essentially loses all predictive power (One can produce whatever
one wants.). However, it would not surprise me if we are forced
soon to consider more complicated models. My guess is that this
may happen when we try to simultaneously fit the synchrotron
and inverse Compton spectra for sources like Mkn 421 where we
may have good, simultaneous, broad-band spectral data.
We may find that the seed photon distribution we need to
produce the observed Compton spectrum given a particular
electron distribution is different from
the synchrotron photon distribution generated by those electrons.
If stellar jets and the microquasars are any guide, jets
are highly episodic and variable phenomena. The current spectrum
we observe may be the superposition of several different flares
or acceleration ``shots''
which are in various stages of cooling down and extinguishing 
themselves. This possibility should not be forgetten when interpreting
data. (Because of ``dilution'' effects,
the variability amplitude at a given energy
may be significantly reduced from what one\break naively expects.)

\vspace{6.5mm}
\leftline{\bf 3.4. What are the ``typical'' bulk Lorentz factors for jets?}
\vspace{4mm}

\noindent 
I note that during the workshop there were several instances 
where jet bulk Lorentz factors as high as $20-40$ were casually tossed
about when talking about emission models  (and Lorentz factors 100+
were invoked to explain away the high brightness temperatures
from intraday variability). These are somewhat higher than what
I was used to. I did not carry out any careful statistics,
but it may be worth double-checking the agreement between emission
model Doppler factors (consistent with rapid variability)
and superluminal motion (VLBI-scale) Doppler factors. The structure of jets
as function of distance from the central source may be more 
complicated and vary more than we currently suspect. The suggestion
by Bicknell (these proceedings) that jets in weak sources start out
as mildly relativistic and then become subrelativistic via entrainment
would be one example of this that definitely deserves more\break{}
investigation.

\vspace{6.5mm}
\leftline{\bf 3.5. What are MeV blazars?}
\vspace{4mm}

\noindent 
Focusing again on high-energy emission from relativistic jets, are there
two distinct classes of gamma-ray emitting blazars: the
``MeV'' blazars (whose energy output peaks strongly in
this energy range) and the more conventional GeV/TeV blazars?
Or is there simply a continuum of blazar spectra corresponding to
(in a Ghisellini-type picture) a range 
of minimum electron injection Lorentz factor and
power law injection indices (see \S 3.2)?
Or do MeV blazars represent a very different, compact and fireball-like
source as argued by Thompson in his talk, or as argued by Sikora, are
they evidence for boosted thermal emission  from some hot, continuously
reheated region? The unification picture is appealing, but some
of the EGRET MeV blazar spectra look rather strange (showing
``lines''?). The jury is still out on this question. 

One further question: the radio galaxy Cen A was detected by GRO
to have variable emission that definitely extended
to beyond 1 MeV in some epochs (Kinzer et al. 1995). Cen A is
not supposed to be a gamma-ray blazar as its jet is 
not pointing at us. However, is the jet responsible for this emission too?
(No Seyferts have been detected at 1 MeV.)

\vspace{6.5mm}
\leftline{\bf 3.6. What is the connection between jets, the accretion disk}
\leftline{\bf\phantom{3.6.} and/or the central black hole?}
\vspace{3mm}

\noindent 
Although this was not addressed directly at the workshop
(but see the contribution from Moderski, Sikora, \& Lasota),
it is still one of the key questions in our quest
to understand the origin of jets. The radio loud vs. radio
quiet quasar (jet vs. no jet) dichotomy has been argued as evidence
for a clear difference in some intrinsic
property of the central objects in these sources.
A popular suggestion is that a jet is produced only when
the black hole is rapidly rotating. However, as seen in 
this workshop, the distinction between radio loud 
and radio quiet is becoming  muddied. Evidence
for outflows that are at least mildly relativistic (in
the initial stages) seems to showing up in most AGN 
when one looks hard enough. In the Galactic camp, we also have
 both confirmed black hole (GRO J1655) and neutron star 
(Cir X-1) binary systems that show jets. The only thing in common
between these systems is presumably the deep gravitational potential well
and the presence of an accretion disk. Perhaps a~jet is mainly
a phenomenon related to the accretion disk, and 
in the case of AGN,  the distinction
between strong and weak radio jet sources has more 
to do with the environmental conditions in
the central region of the AGN (e.g., the density of ambient
gas) rather than black hole?  (For example if a starburst
is going on in the center, the central gas density may be very high
and entrainment will ``kill off'' the jet.)
Motivated by detailed numerical simulations, Meier et al. (1997) have
also proposed a very interesting picture where all disks generate
outflows with a  kinetic luminosity that grows as the
characteristic strength of the magnetic field in the disk grows
(with the strength of the field in the disk presumably scaling
with the mass of the central object). The outflow is mildly
relativistic until the disk field strength exceeds a critical
``switch'' and the flow then becomes strong relativistic.
This could explain the apparent continuity in FR I/FR II radio
power, but the relatively sharp distinction in FR I/FR II 
radio morphology. On the other hand, based on the relative
intensities of the synchrotron continuum and broad emission
lines, BL Lac/FR I galaxies
have disks that appear to be  subluminous compared to those in radio loud
galaxies. To power the jet, we may require an additional source
of energy, such as could be provided by the central black hole
via the Blandford-Znajek mechanism (In principle, then,
when the fueling of the black hole has almost completely
stopped, we could still see a jet.).

\vspace{10.5mm}
\leftline{\large\bf 4. Future prospects}
\vspace{7mm}

\tolerance200 
\noindent 
As breathtaking as it has been, the rush of observational information
on jets is likely to continue. On the radio front, we have only
just begun to exploit the capabilities of the VLBA. For the first time we may
able to discern ambient gas clouds illuminated by the high brightness
temperature emission of a jet and thus learn more about 
the environment through which the jet propagates. Space based VLBI,
such as is already being carried out using the VSOP satellite, will further
increase the spatial resolution available to us and will either
resolve the cores of radio-loud AGN or raise the lower limits on the 
brightness 
temperature of cores from the already uncomfortably high values of 
$\sim 10^{12}K$ observed in some cases (At such high brightness temperatures,
induced Compton and Raman scattering effects may become important,
e.g., see Coppi, Blandford \& Rees 1993.). 
Unlike the IDV brightness temperature estimates based on
variability, these are based on direct spatial 
constraints and are much more robust. In general, the VLBA together
with an upgraded European VLBI network will allow for much more frequent
monitoring of  many more sources than has been possible before.
Multi-frequency polarization observations (including perhaps
circular polarization) should become relatively routine.
Gabuzda has already presented an example here where  
monitoring of PKS 2155 found a 5 hr (intraday) variation in the polarization
of one VLBI component but no variations in another. Such
behavior is not entirely unexpected as the result of interstellar
scintillation, but if confirmed in other sources, 
it may lead to some interesting constraints on 
the interstellar scintillation explanation for IDV.

\tolerance500 
At optical wavelengths, continued use of HST coupled with high 
resolution observations of the VLA  will provide us 
more detailed examples of jet-ISM interactions, such as we have begun to 
see at this conference. Combined high spatial resolution optical
and radio observations will also allow  us to watch how a 
synchrotron-emitting population
of electrons ages and provides constraints on the particle acceleration
mechanisms, e.g., as was done for the shocks in the jet of M87 
(Stiavelli et al., 1997). In addition, several large CCD mosaic instruments
will be coming on-line in the next few years (e.g., the Sloan
Digital Sky Survey, Megacam at CFHT). These will allow 
deep quasar surveys covering much of the sky. Over the next few years, 
the number of optically
confirmed quasars, including those with jets, should increase at all redshifts
($z<5$) by well over an
order of magnitude. The current world sample of quasars is of order
10,000; the 2DF survey in Australia has already obtained a list of 
25,000 good UV-selected candidates and will take spectra of all them
in the next few years. The optical surveys can be cross-correlated
with radio and X-ray surveys, producing a very large, multi-wavelength
sample which will enable us to test, for example,  
unification schemes and orientation effects in even more detail.
\vspace{1pt}

In the X-ray range, the arrival of AXAF with its 0.5'' spatial resolution
and excellent spectral resolution will allow us to map in more detail
X-ray emission from the outer regions of the jet (e.g., see Diana
Worral in this proceedings) or study how a jet shocks and interacts
with an intracluster medium (e.g., as in the case of NGC 1275, 
Fabian et al. 1994). Of interest to those modeling high energy 
emission from jets, the arrival of satellites with broad-band 
($\sim 1 keV - 100$ keV) capabilities
and good sensitivity like SAX and XTE will allow us to {\it simultaneously\/}
monitor the time evolution of the synchrotron 
and Compton emission components (assuming the SSC picture is 
correct). In one shot, we can monitor the evolution
of the emitting particle distribution at the highest
{\it and\/} the lowest energies (the high energy electrons produce 
the typically $\sim $ keV synchrotron emission and the low energy electrons
are responsible for the Compton upscattered emission 
observe at higher photon energies).
Such information will be crucial for testing inhomogeneous vs. homogeneous jet
emission models and detailed acceleration scenarios. XTE will be particularly
useful because of its large collecting area. In the case of Mkn 421, TeV
flares were observed to occur on timescales as short as half an hour
(Gaidos et al. 1994). With Mkn 421 at its typical emission level, XTE should
 be able  to probe variability down to the level of a few minutes. 
The expectation is that we will finally reach the shortest
timescales on which blazars vary and effects due to finite source
size and finite acceleration time become apparent (Perhaps we will see
the ``reverse hysteresis'' discussed by Kirk in his contribution.).
If we do not see evidence for a lower limit on variability 
timescales, then the corresponding limits on the size of the emitting
region and the jet Doppler beaming factors become even more interesting.
(Already, in Mkn 421 beaming factors $\delta \sim 15$ are talked
about, which is considerably larger than typical BL Lac beaming factor
$\sim 4$ obtained from unification scheme and superluminal
motion  studies.) 
Instruments like XTE and SAX also typically
include all-sky monitors designed to pick up transient events,
such as the flare of a low mass X-ray binary system in our galaxy.
The number of known galactic  black hole binary systems, including
``microquasars'', should thus increase significantly over the 
next few years.

In the range 100 keV - few MeV, which has traditionally been 
difficult to observe, significant improvements in sensitivity
compared to GRO should come from the planned launches of 
the Integral and Astro-E satellites (Current data in this 
range comes mainly
from the Comptel instrument on GRO, which has a~rather low 
sensitivity.). This is the energy range
where the high-energy spectra of blazars show a~break, resulting
perhaps from the inefficient cooling or escape of electrons/pairs below
a~certain energy. In some cases, the ``MeV blazars'', the 
break actually appears to be very sharp and the energy output of 
the blazar peaks in this energy range. 
The explanation for this break is not at all
clear. As discussed by Sikora and Thompson here, 
 understanding it has important implications for the blazar
emission model as~a~whole. 

Finally in the range of 10 MeV and above, we will have a temporary
dearth of data now that the high-energy EGRET instrument on GRO
has come to the end of its useful life and  a replacement for EGRET,
perhaps GLAST, is not likely to be launched for at least
several (ten) years. This deficit in energy coverage, however,
is being rapidly filled in by ground-based Cherenkov detectors.
The  low-energy threshold of these detectors is currently 
$\sim 200$ GeV, but work is already underway or planned 
(e.g. the MAGIC, CELESTE, STACEE)  to lower
this threshold to $\sim 50$ GeV.  The currently existing Cherenkov
detectors (Whipple, HEGRA, CAT, CANGAROO),  however, have  already proven 
themselves to be extremely useful for nearby blazars like Mkn 421
and Mkn 501 (e.g., see the papers on the recent Mkn 501 flare by
Bradbury et al. 1997, and Catanese et al. 1997). Perhaps the key attribute
of these detectors to bear in mind is their very large collecting
area, already at least $10^4$ times that of EGRET. Since these
detectors are located on the ground, they are not subject to the
stringent size and weight limitations that constrain
space-based detectors 
(The sensitivity of a~Cherenkov detector can in principle
be increased arbitrarily by putting together an ever larger array
of Cherenkov mirrors.). The result is that while EGRET can resolve
flares down to timescales of several hours (for the very brightest
events), current Cherenkov detectors  can probe variability
on timescales of 15-30 minutes. Given the extreme variability
of blazars, such timing capability is crucial and complements well
the capabilities of an instrument like XTE. 
 Detailed correlation
studies of X-ray and TeV variability will be among the most important
in constraining and testing the SSC model for the emission of 
nearby, weak blazars like Mkn 421. (In fact, successful studies of this
type have already been carried using ASCA and Whipple, e.g., 
see the contribution of Takahashi et al.)
With stereoscopic imaging techniques, Cherenkov  energy resolutions 
$\sim 25\%$ can be achieved at $\sim$ TeV energies. Thus it will
be possible to carry out detailed spectral, not just intensity, comparisons.

From my perspective, the next few years should prove rather exciting.
I look forward to the next meeting in Krak\'ow...

\vspace{8mm}

\leftline{\large\bf Acknowledgments}
\vspace{5mm}

\noindent 
I wish to thank the organizers for their patience and for
all their hard work in putting together a smoothly run and
memorable workshop.

\vspace{8mm}
 \leftline{\large\bf References}
\vspace{5mm}

\references

Baum, S. et al, 1997, ApJ, 483, 178

Begelman, M.C., Rees, M.J., Sikora, M. 1994, ApJL, 429, 57

Bicknell, G.V., Dopita, A.M., Odea, C.P.O 1997, ApJ, 485, 112

Bjornsson, C.I. 1990, MNRAS, 242, 158

Bradbury, S. et al. 1997, A\&A, 320, 5 

Catanese, M. et al. 1997, ApJL, 487, 143

Coppi, P.S., Aharonian, F.A. 1997, ApJL, 487, 9

Coppi P.S., Blandford R.D., Rees M.J. 1993, MNRAS, 262, 603
   
Gaidos, J.A. et al., 1996, Nat, 383, 319

Herrnstein, J.R. et al. 1997, ApJL, 475, 17

Homan, D.C., Wardle, J.F.C., Roberts, D.H. 1997, in {\it IAU
Colloqium 164: Radio Emission from Galactic and Extragalactic
Compact Sources}, in press

Junor, W., Biretta, J.A. 1995, AJ, 109, 501

Kedziora-Chudczer, L. et al. 1996, IAUC 6418 

Kinzer, J. et al. 1995, ApJ, 449, 105

Lind, K.R., Blandford, R.D. 1985, ApJ, 295, 358

Mattox, J.R. et al. 1997, ApJ, 476, 692

Meier, D.L. et al. 1997, Nat, 388, 350

Melia, F., K\"onigl, A. 1989, ApJ, 340, 162

Padovani, P., Giommi, P., Fiore, F. 1997, MNRAS, 284, 569 

Pohl, M. et al. 1995, A\&A, 303, 383

Pohl, M. et al. 1997, A\&A, 326, 51

Rawlings, S.G., Saunders, R.D.E, 1991, Nat, 349, 138

Sikora, M., Sol, H., Begelman, M.C., Madejski, G.M., 1996, MNRAS, 280, 781

Stiavelli, M. et al. 1997, MNRAS, 285, 181

Unwin, S. et al. 1997, ApJ, 480, 596

Urry, C.M., Padovani, P. 1995, PASP, 107, 803

Wehrle, A.E. et al. 1993, BAAS, 183, \#107

Zensus, A. et al. 1997, ApJ, 443, 35

\end{document}